**О. В. Дудник** [1, 2], **Є. В. Курбатов** [1]

[1] Радіоастрономічний інститут Національної академії наук України, Харків, Україна
[2] Харківський національний університет ім. В. Н. Каразіна, Харків, Україна

# ВИКОРИСТАННЯ НАНОСУПУТНИКІВ ДЛЯ ВИВЧЕННЯ ПРИРОДИ МІКРОСПЛЕСКІВ ВИСОКОЕНЕРГЕТИЧНИХ ЧАСТИНОК У МАГНІТОСФЕРІ ЗЕМЛІ: ІДЕЯ КОСМІЧНОГО ЕКСПЕРИМЕНТУ

*Представлено концепцію наукового космічного експерименту з вивчення природи мікросплесків заряджених частинок високих енергій у магнітосфері Землі з використанням наносупутникової платформи у форматі «CubeSat». Дано опис функціональної схеми, структурних одиниць і технічних характеристик мініатюрного реєстратора-аналізатора електронів і протонів MiPA_ep.*

***Ключові слова:*** *радіаційний пояс, електрон, наносупутник, «CubeSat», кремнієвий PIN-детектор, органічний сцинтилятор, Бразильська магнітна аномалія, нахил орбіти супутника.*

## ВСТУП

У земній магнітосфері є шари, завжди заповнені високоенергетичними електронами і протонами. Вони утворюють радіаційні пояси Землі, або так звані пояси Ван Аллена. Кількість експериментальних даних про іонізовану радіацію довкола Землі безперервно зростає. Поновлення інтересу в останні роки до вивчення радіаційних поясів пов'язане з розвитком принципово нових технологій, як у детекторах для реєстрації елементарних заряджених частинок, так і в космічній мікроелектроніці. Крім відомих поясів Ван Аллена, досліджених за допомогою великої кількості штучних супутників Землі за часи космічної ери, є просторові області, що спорадично заповнюються низькоенергетичними і субрелятивістськими електронами. Одним зі значних наукових досягнень у вивченні іонізованої радіації є виявлення мікросплесків високоенергетичних електронів, які реєструються на краях зовнішнього електронного поясу [2, 10, 12, 13], на низьких широтах та поблизу екваторіальної зони [9, 11, 15]. Інший напрям досліджень пов'язаний з реєстрацією короткоживучих радіаційних поясів у проміжку між внутрішнім і зовнішнім поясами Ван Аллена, а також під внутрішнім електронним поясом [1, 8].

Зони радіації, що спорадично генеруються у проміжку між поясами Ван Аллена, на середніх широтах, мікросплески частинок на краях радіаційних поясів не є ще достатньо вивченими. Мапа розподілу електронних потоків на висоті $h \approx 600$ км показує, що частинки розподілено приблизно у трьох широтних зонах: 1) у екваторіальній зоні (параметр Мак-Ілвайна $L < 1.2$), 2) у низькоширотній зоні ($1.2 < L < 1.4$), 3) на середніх широтах ($1.6 < L < 2$) [15]. Однак дослідники до цього часу не запропонували механізмів генерації електронних піків на середніх та низьких широтах, що спостерігаються на низькоорбітальних супутниках. Крім того, в літературі не-має згадувань про існування вузького довгоживучого прошарку зарядженої радіації високої енергії під внутрішнім радіаційним поясом Землі, що вперше був задекларований на основі спостережень приладом СТЕП-Ф на борту космічного апарата «КОРОНАС-Фотон» [4].

Ще однією актуальною і важливою для людства проблемою є визначення провісників сейсмічної активності через реєстрацію висипань частинок з радіаційних поясів в атмосферу Землі перед землетрусами [5—7, 16, 18]. Базуючись на 11-річних даних з 1998 по 2011 рр., зібраних за допомогою приладу MEPED на супутнику NOAA-15/POES щодо потоків електронів і протонів на висоті близько 800 км, статистично суттєва кореляція була знайдена між збуреннями потоку електронів з енергіями $E_e$ = 30...100 кеВ та магнітудою землетрусів ($M \geq 6$), що мали місце за цей період в регіоні острова Суматра та Філіппін [5]. При цьому електрони, що висипались на малих $L$-оболонках, передували головному поштовху землетрусу на 2—3 год. В роботі [5] зроблено висновок, що для реєстрації сплесків електронів перед землетрусами детекторні системи мають бути чутливими до електронів з енергіями до $E_e \sim 1$ МеВ, вісь кута зору має бути спрямована вертикально, тоді як сам кут зору, сформований коліматорною системою, повинен бути не меншим за $\Delta\theta = 30°$. Орбіта супутника, яка максимально збільшує площу детектування електронів, що висипаються з радіаційних поясів, — це екваторіальна орбіта з висотами від 900 до 1000 км. Крім того, подальші дослідження можуть бути більш плідними, якщо залучити до експерименту інші супутникові вимірювання, беручи до уваги денно-нічний ефект у розподілах частинок на низьких орбітах.

З метою вирішення перелічених наукових задач в теперішній час розробляються прилади різного ґатунку. Зокрема, іноземними інженерами плануються або вже здійснені впровадження мініатюрних супутникових приладів елементарних заряджених частинок високих енергій. Індійськими вченими і студентами з Інституту технології міста Мадрас розроблено компактний прилад SPEED для реєстрації електронів і протонів на борту наносупутника IITMSAT [17]. IITMSAT призначений для вивчення висипань заряджених частинок з радіаційних поясів Ван Аллена, пов'язаних з сейсмо-електромагнітними емісіями. В іншому приладі RADMON студентського наносупутника «Aalto-1» (Фінляндія) як детектори використовуються кремнієвий PIN-детектор і сцинтиляційний детектор на основі монокристала CsI(Tl) з кремнієвим фотодіодом великої площі [14]. Час світлового спалаху сцинтилятора CsI(Tl) досить великий ($\tau > 3$ мкс), що не дає змоги реєструвати потоки частинок зі значною щільністю. У приладі американської місії FIREBIRD у форматі «CubeSat 1.5 U» прилади вимірювали на низьких орбітах сплески електронів у п'яти діапазонах парами іонно-імплантованих кремнієвих детекторів у полі зору, що складав приблизно $\Delta\theta \approx 45°$ [3].

В роботі представлена ідея здійснення супутникового експерименту з дослідження природи мікросплесків високоенергетичних заряджених частинок з використанням компактного приладу, відмінними рисами якого є: а) застосування швидкодіючого сцинтиляційного детектора з монокристала паратерфенілу вітчизняної розробки, нечутливого до вторинного гальмівного випромінювання; б) реєстрація частинок з двох протилежних напрямів, що дозволить вивчати ступінь анізотропії мікросплесків.

### НАУКОВІ ЗАДАЧІ СУПУТНИКОВОГО ЕКСПЕРИМЕНТУ

Метою здійснення супутникового експерименту є вивчення механізмів систематичних та спорадичних варіацій та сплесків інтенсивності високоенергетичних електронів та протонів на низьких орбітах. Задачами досліджень є: а) перевірка факту існування додаткового внутрішнього електронного радіаційного поясу на $L \approx 1.6$ для частинок з енергіями від десятків кеВ до $E_e \approx 0.5$ МеВ у геомагнітно спокійних умовах; б) визначення енергетичних спектрів частинок у стаціонарних радіаційних поясах та у мікросплесках поза межами поясів і аналіз тонкої структури мікросплесків; в) вивчення ступеня анізотропії напрямів швидкостей електронів всередині радіаційних поясів та у мікросплесках на краях поясів Ван Аллена та поза їхніми межами під час проявів со-

О. В. Дудник, Є. В. Курбатов

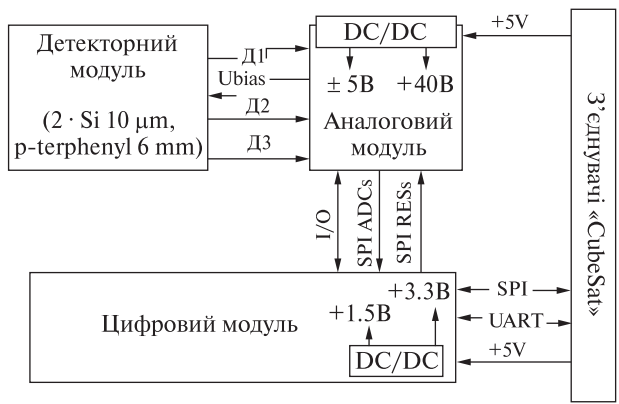

*Рис. 1.* Функціональна схема приладу

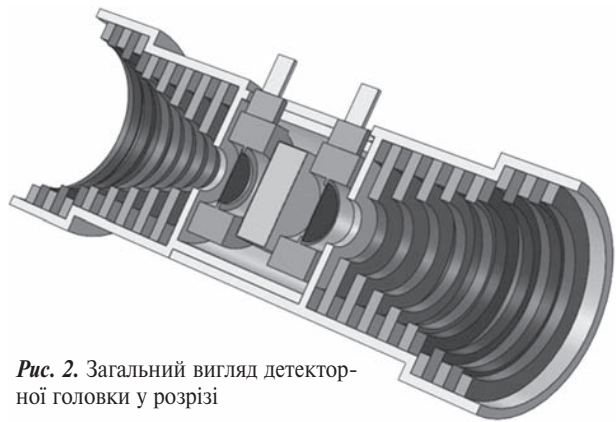

*Рис. 2.* Загальний вигляд детекторної головки у розрізі нячної, магнітосферної та іоносферної активностей; г) уточнення та визначення часу життя, коефіцієнтів радіальної та пітч-кутової дифузії електронів у зовнішніх та внутрішніх шарах магнітосфери Землі в умовах низької, помірної та екстремальної геомагнітної активності; д) пошук і визначення відмінних рис поміж мікросплесками електронів, породжених магнітосферою, сонячною і міжпланетною активністю, та таких, що виникли в результаті сейсмічної активності.

Перелічені задачі планується вирішувати за допомогою компактного супутникового приладу власної розробки. З цією метою пропонується розробити, виготовити та впровадити прилад MiPA_ep (Мініатюрний Реєстратор-Аналізатор електронів і протонів), що буде накопичувати дані про потоки частинок в різних діапазонах енергій на різних широтах та довготах, у різні періоди сонячної та геомагнітної активності, на освітленій та затемненій сторонах земної поверхні.

### ПРИНЦИПИ ПОБУДОВИ МІНІАТЮРНОГО ПРИЛАДУ MiPA_ep

*Блок-схема та основні параметри приладу.* Як корисне навантаження наносупутника, побудованого у форматі «CubeSat», прилад MiPA_ep складатиметься з трьох модулів: детекторного, аналогового та цифрового (рис. 1). Кожен з модулів є окремою одиницею, що з'єднуються за допомогою кабелів та з'єднувачів. У аналоговому та цифровому модулях будуть встановлені перетворювачі напруги, що забезпечуватимуть відповідні блоки необхідними рівнями напруги.

Нижче представлено основні наукові та технічні параметри приладу:

| | |
|---|---|
| Кут зору телескопічної системи, град | 34 |
| Геометричний фактор, см$^2$ср | 0.76 |
| Сорти та діапазони енергій заряджених частинок, MeV: | |
| електрони | 0.04...2.5 |
| протони | 1.25...4.7 |
| Кількість напрямків реєстрації (0...180°) | 2 |
| Кількість енергетичних каналів кожного сорту частинок (на кожний напрямок) | 5 |
| Кількість енергетичних каналів без визначення сорту частинок (на кожний напрямок) | 5 |
| Мінімальна часова роздільна здатність потоків частинок, с | 0.1 |
| Інформативність приладу (середня 6370 біт/с), Мбайт/добу | 66 |
| Об'єм цифрового масиву з науковими даними, байт | 796 |
| Період формування інформаційних масивів для надсилання до системи збору інформації, с | 1 |
| Кількість програмних команд для керування приладом | 6 |
| Розмір програмної команди, байт | 12 |
| Середня частота надсилання команд, на добу | 4 |
| Період отримання часових міток від СЗІ, с (точність прив'язки до бортового часу не гірша за 1 мс) | 1 |
| Період отримання даних про орієнтацію приладу, с | 1 |
| Розміри модулів, мм: | |
| детекторний | ⌀ 34 × 100 |
| аналоговий | 90.2 × 95.9 × × 23.1 |
| цифровий | 90.2 × 95.9 × × 23.1 |
| Вага, г | < 850 |

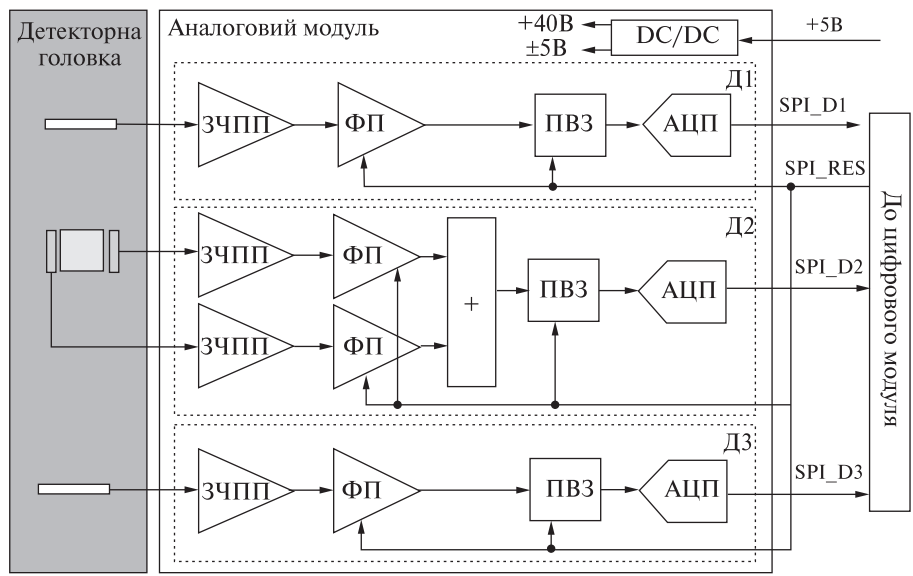

***Рис. 3.*** Функціональна схема аналогового модуля з підключеним детекторним модулем

Споживана потужність, Вт:
   середня ............................. 0.7
   пікова .............................. 1
Напруга живлення, В................. +5 ± 10 %
Струм живлення, А:
   середній ............................ 0.14
   піковий ............................. 0.2

***Детекторний модуль.*** Детекторний модуль (ДМ) складатиметься з двох кремнієвих «прольотних» детекторів і одного сцинтиляційного детектора на основі монокристала паратерфенілу, що виконуватиме функцію детектора повного поглинання. Модуль буде містити також коліматорну систему, що формуватиме кут зору у двох протилежних напрямках (рис. 2). Детектори розміщуються у механічній конструкції детекторного модуля таким чином, щоб утворювалася двонаправлена телескопічна система. Коліматорна система формує кут зору приладу $\Delta\theta = 34°$.

***Аналоговий і цифровий модулі.*** Аналоговий модуль складатиметься з трьох спектрометричних каналів: двох ідентичних каналів для обробки сигналів з кремнієвих детекторів Д1 та Д2 і одного каналу для підключення каналу сцинтиляційного детектора Д3. Канали Д1 та Д3 складаються з зарядочутливого попереднього підсилювача (ЗЧПП), формувального підсилювача, пристрою вибірки та запам'ятовування та 12-розрядного АЦП. Канал Д2 має два ЗЧПП для підключення фотодіодів, два формувальних підсилювачі, сигнали з виходів яких надходять до аналогового суматора, пристрою вибірки та запам'ятовування та 12-розрядного АЦП. Функціональну схему аналогового модуля з підключеним детекторним модулем наведено на рис. 3.

Аналоговий модуль матиме окремі цифрові SPI-інтерфейси для зчитування кодів АЦП кожного із спектрометричних каналів та один загальний SPI-інтерфейс для керування параметрами каналів. Планується регулювати коефіцієнти підсилення формувачів та рівні спрацювання пікових детекторів. Модуль також містить в собі перетворювачі вхідної напруги у необхідні рівні живлення.

Основними задачами цифрового модуля є керування параметрами аналогового модуля; зчитування та опрацювання кодів АЦП аналогового модуля; визначення сортів та енергій зареєстрованих частинок; формування вихідного телеметричного кадру; накопичення та вивантаження наукових даних; прийом та відпрацювання командних повідомлень.

## ВИМОГИ ДО РОЗМІЩЕННЯ ПРИЛАДУ ТА КОНФІГУРАЦІЯ ОРБІТИ

Нахилення орбіти супутника, що перекриває $L$-оболонки від $L \approx 1.0$ (геомагнітний екватор) до $L \approx 2.5$ (зовнішній край внутрішнього радіацій-ного поясу Ван Аллена) на малих висотах має дорівнювати φ ≈ 50°. Якщо у процесі виконання космічної місії ставити задачу одночасно спостерігати за динамікою частинок у зазорі між двома поясами Ван Аллена та у зовнішньому радіаційному поясі, то нахилення орбіти має бути не меншим, ніж φ ≈ 75°. Виконання задачі моніторингу тільки мікросплесків електронів під радіаційними поясами обмежує величину нахилення орбіти до φ ≈ 20...30°. Але і в цьому разі можна буде слідкувати за потоками електронів у внутрішньому радіаційному поясі через перетинання космічним апаратом Бразильської магнітної аномалії. Попередні експерименти показали, що висоти орбіти супутників, які максимально збільшують площу детектування електронів, що висипаються з радіаційних поясів, становлять від 900 до 1000 км.

Кут зору детекторної головки приладу має бути спрямований вертикально вгору по відношенню до поверхні Землі з метою реєстрації саме потоків частинок, що висипаються. В такому разі з другого (протилежного) напряму прилад буде здатний реєструвати високоенергетичні електрони і протони, що рухаються від атмосфери та іоносфери вертикально вгору. Таким чином, з'явиться можливість вивчення ступеня анізотропії потоків частинок, пошуку зв'язку з земними гама-сплесками (Terrestrial Gamma Flashes) тощо.

## ВИСНОВКИ

Експериментальні дослідження динаміки потоків частинок в радіаційних поясах Землі та поза їхніми межами, вивчення природи мікросплесків високоенергетичних заряджених частинок доцільно здійснити за допомогою мініатюрного економічного реєстратора-аналізатора MiPA_ep, відмінними рисами якого є застосування легкого органічного сцинтилятора з монокристала паратерфенілу та реєстрація частинок з двох протилежних напрямів, що дозволить вивчати ступінь анізотропії мікросплесків. Невеликі геометричні розміри модулів, малі вага ($m < 850$ г), споживана потужність ($P < 0.7$ Вт) і первинна напруга живлення ($U = 5$ В) надають змогу застосувати прилад MiPA_ep як корисне навантаження студентських наносупутників у форматі «CubeSat».

ЛІТЕРАТУРА

1. *Baker D. N., Kanekal S. G., Hoxie V. C., Henderson M. G., Li X., Spence H. E., et al.* A Long-Lived Relativistic Electron Storage Ring Embedded in Earth's Outer Van Allen Belt // Science. — 28 February 2013. — P. 1—7.
2. *Blake J. B., Looper M. D., Baker D. N., Nakamura R., Klecker B., Hovestadt D.* New high temporal and spatial resolution measurements by SAMPEX of the precipitation of relativistic electrons // Adv. Space Res. — 1996. — **18**, N 8. — P. 171—186.
3. *Crew A. B., Spence H. E., Blake J. B., Klumpar D. M., Larsen B. A., et al.* First multipoint in situ observations of electron microbursts: Initial results from the NSF FIREBIRD II mission // J. Geophys. Res., Space Phys. — 2016. — **121**, N 6. — P. 5272—5283.
4. *Dudnik O. V., Podgórski P., Sylwester J., Gburek S., Kowaliński M., et al.* Investigation of Electron Belts in the Earth's Magnetosphere with the Help of X-ray Spectrophotometer SphinX and Satellite Telescope of Electrons and Protons STEP-F: Preliminary Results // Space Sci. and Technology. — 2011. — **17**, N 4. — P. 14—25.
5. *Fidani C.* Particle precipitation prior to large Earthquakes of both the Sumatra and Philippine Regions: a statistical analysis // J. Asian Earth Sci. — 2015. — **114**, Part 2. — P. 384—392.
6. *Fidani C., Battiston R.* Analysis of NOAA particle data and correlations to seismic activity // Natur. Hazards and Earth System Sci. — 2008. — **8**, N 6. — P. 1277—1291.
7. *Fidani C., Battiston R., Burger W. J.* A study of the correlation between Earthquakes and NOAA satellite energetic particle bursts // Remote Sens. — 2010. — **2**, N 9. — P. 2170—2184.
8. *Hudson M. K., Kress B. T., Mueller H.-R., Zastrow J. A., Blake J. B.* Relationship of the Van Allen radiation belts to solar wind drivers // J. Atmos. and Solar-Terr. Phys. — 2008. — **70**, N 5. — P. 708—729.
9. *Imhof W. L., Reagan J. B., Gaines E. E.* The energy selective precipitation of inner zone electrons // J. Geophys. Res. — 1978. — **83**, N A9. — P. 4245—4254.

10. *Imhof W. L., Voss H. D., Mobilia J., Datlowe D. W., Gaines E. E., et al.* Relativistic Electron Microbursts // J. Geophys. Res. — 1992. — **97**, N A9. — P. 13829—13837.
11. *Nagata K., Kohno T., Murakami H., Nakamoto A., Hasebe N., et al.* Electron (0.19—3.2 MeV) and proton (0.58—35 MeV) precipitations observed by ONZORA satellite at low latitude zones $L = 1.6—1.8$ // Planet. and Space Sci. — 1988. — **36**, N 6. — P. 591—606.
12. *Nakamura R., Baker D. N., Blake J. B., Kanekal S., Klecker B., Hovestadt D.* Relativistic electron precipitation enhancements near the outer edge of the radiation belt // Geophys. Res. Lett. — 1995. — **22**, N 9. — P. 1129—1132.
13. *Nakamura R., Isowa M., Kamide Y., Baker D. N., Blake J. B., Looper M.* SAMPEX observations of precipitation bursts in the outer radiation belt // J. Geophys. Res. — 2000. — **105**, N A7. — P. 15875—15885.
14. *Peltonen J., Hedman H-P., Ilmanen A., Lindroos M., Maattanen M., et. al.* Electronics for the RADMON instrument on the Aalto-1 student satellite // 10[th] European Workshop on Microelectronics Education (EWME). (14—16 May 2014, Tallinn, Estonia). — Proceedings. — P. 161—166.
15. *Sadovnichy V. A., Panasyuk M. I., Yashin I. V., Barinova V. O., Veden'kin N. N., et. al.* Investigations of the Space Environment Aboard the Universitetsky—Tat'yana and Universitetsky—Tat'yana-2 Microsatellites // Solar System Res. — 2011. — **45**, N 1. — P. 3—29.
16. *Sgrigna V., Carota L., Conti L., Corsi M., Galper A. M., et al.* Correlations between earthquakes and anomalous particle bursts from SAMPEX/PET satellite observations // J. Atmos. and Solar-Terr. Phys. — 2005. — **67**, N 15. — P. 1448—1462.
17. *Surya Teja S. S., Subramanyan V., Elangovan R., Reddy L. M., Ramachandran D., et al.* Design of Nuclear Instrumentation for Space-based Proton-Electron Energy Detector (SPEED) // International Conference on Space Science and Communication (IconSpace) (10—12 August 2015, Kuala Lumpur, Malaysia): Proceedings. — P. 181—186.
18. *Zhang X., Fidani C., Huang J., Shen X., Zerren Z., Qian J.* Burst increases of precipitating electrons recorded by the DEMETER satellite before strong earthquakes // Natur. Hazards and Earth System Sci. — 2013. — **13**, N 1. — P. 197—209.

REFERENCES

1. Baker D. N., Kanekal S. G., Hoxie V. C., Henderson M. G., Li X., Spence H. E., et al. A Long-Lived Relativistic Electron Storage Ring Embedded in Earth's Outer Van Allen Belt. *Science*. 1—7 (28 February 2013).
2. Blake J. B., Looper M. D., Baker D. N., Nakamura R., Klecker B., Hovestadt D. New high temporal and spatial resolution measurements by SAMPEX of the precipitation of relativistic electrons. *Advances in Space Research*. **18** (8), 171—186 (1996).
3. Crew A. B., Spence H. E., Blake J. B., Klumpar D. M., Larsen B. A., et al. First multipoint in situ observations of electron microbursts: Initial results from the NSF FIREBIRD II mission. *Journal of Geophysical Research, Space Physics*. **121** (6), 5272—5283 (2016).
4. Dudnik O. V., Podgórski P., Sylwester J., Gburek S., Kowaliński M., et al. Investigation of Electron Belts in the Earth's Magnetosphere with the Help of X-ray Spectrophotometer SphinX and Satellite Telescope of Electrons and Protons STEP-F: Preliminary Results. *Space Science and Technology*. **17** (4), 14—25 (2011).
5. Fidani C. Particle precipitation prior to large Earthquakes of both the Sumatra and Philippine Regions: a statistical analysis. *Journal of Asian Earth Sciences*, **114** (2), 384—392 (2015).
6. Fidani C., Battiston R. Analysis of NOAA particle data and correlations to seismic activity. *Natural Hazards and Earth System Sciences*. **8** (6), 1277—1291 (2008).
7. Fidani C., Battiston R., Burger W. J. A study of the correlation between Earthquakes and NOAA satellite energetic particle bursts. *Remote Sensing*, **2** (9), 2170—2184 (2010).
8. Hudson M. K., Kress B. T., Mueller H.-R., Zastrow J. A., Blake J. B. Relationship of the Van Allen radiation belts to solar wind drivers. *Journal of Atmospheric and Solar-Terrestrial Physics*. **70** (5), 708—729 (2008).
9. Imhof W. L., Reagan J. B., Gaines E. E.. The energy selective precipitation of inner zone electrons. *Journal of Geophysical Research*. **83** (A9), 4245—4254 (1978).
10. Imhof W. L., Voss H. D., Mobilia J., Datlowe D. W., Gaines E. E., et al. Relativistic Electron Microbursts. *Journal of Geophysical Research*. **97** (A9), 13829—13837 (1992).
11. Nagata K., Kohno T., Murakami H., Nakamoto A., Hasebe N., et al. Electron (0.19—3.2 MeV) and proton (0.58—35 MeV) precipitations observed by ONZORA satellite at low latitude zones $L = 1.6—1.8$. *Planetary and Space Science*. **36** (6), 591—606 (1988).
12. Nakamura R., Baker D. N., Blake J. B., Kanekal S., Klecker B., Hovestadt D. Relativistic electron precipitation enhancements near the outer edge of the radiation belt. *Geophysical Research Letters*. **22** (9), 1129—1132 (1995).
13. Nakamura R., Isowa M., Kamide Y., Baker D. N., Blake J. B., Looper M. SAMPEX observations of precipitation bursts in the outer radiation belt. *J. Geophys. Res.* **105** (A7), 15875—15885 (2000).
14. Peltonen J., Hedman H-P., Ilmanen A., Lindroos M., Maattanen M., et al. Electronics for the RADMON instrument on the Aalto-1 student satellite. *10[th] European Workshop on Microelectronics Education (EWME), 14—16 May 2014*, Tallinn, Estonia. Proceedings. P. 161—166 (2014).

*О. В. Дудник, Є. В. Курбатов*

15. Sadovnichy V. A., Panasyuk M. I., Yashin I. V., Barinova V. O., Veden'kin N. N., et al. Investigations of the Space Environment Aboard the Universitetsky—Tat'yana and Universitetsky—Tat'yana-2 Microsatellites. *Solar System Research*. **45** (1), 3—29 (2011).
16. Sgrigna V., Carota L., Conti L., Corsi M., Galper A. M., et al. Correlations between earthquakes and anomalous particle bursts from SAMPEX/PET satellite observations. *J. Atmospheric and Solar-Terrestrial Physics*. **67** (15), 1448—1462 (2005).
17. Surya Teja S. S., Subramanyan V., Elangovan R., Reddy L. M., Ramachandran D., et al. Design of Nuclear Instrumentation for Space-based Proton-Electron Energy Detector (SPEED). *International Conference on Space Science and Communication (IconSpace) (10—12 August 2015, Kuala Lumpur, Malaysia*. Proceedings, 181—186 (2015).
18. Zhang X., Fidani C., Huang J., Shen X., Zerren Z., Qian J. Burst increases of precipitating electrons recorded by the DEMETER satellite before strong earthquakes. *Natural Hazards and earth System Sciences*. **13** (1), 197—209 (2013).

*А. В. Дудник* [1,2], *Е. В. Курбатов* [1]

[1] Радиоастрономический институт Национальной академии наук Украины, Харьков, Украина
[2] Харьковский национальный университет им. В. Н. Каразина, Харьков, Украина

ИСПОЛЬЗОВАНИЕ НАНОСПУТНИКОВ ДЛЯ ИЗУЧЕНИЯ ПРИРОДЫ МИКРОВСПЛЕСКОВ ВЫСОКОЭНЕРГЕТИЧЕСКИХ ЧАСТИЦ В МАГНИТОСФЕРЕ ЗЕМЛИ: ИДЕЯ КОСМИЧЕСКОГО ЭКСПЕРИМЕНТА

Представлена концепция научного космического эксперимента по изучению природы микровсплесков заряженных частиц высоких энергий в магнитосфере Земли с использованием наноспутниковой платформы в формате «CubeSat». Дано описание функциональной схемы, структурных единиц и технических характеристик миниатюрного регистратора-анализатора электронов и протонов МиРА_ep.

*Ключевые слова*: радиационный пояс, электрон, наноспутник, кремниевый детектор, органический сцинтиллятор, Бразильская магнитная аномалия, наклонение орбиты спутника.

*O. V. Dudnik* [1,2], *E. V. Kurbatov* [1]

[1] Institute of Radio Astronomy, NAS of Ukraine, Kharkiv, Ukraine
[2] V. N. Karazin Kharkiv National University, Kharkiv, Ukraine

NANOSATELLITES FOR THE STUDY OF HIGH ENERGY PARTICLES' MICROBURSTS' NATURE IN THE EARTH'S MAGNETOSPHERE: AN IDEA OF COSMIC EXPERIMENT

A concept of a cosmic scientific experiment is presented. The main goal of the experiment is the study of microbursts of charged particles of high energy in the Earth's magnetosphere. The experiment is designed to use a nanosatellite platform. The paper describes the functional scheme, structural features and technical characteristics of a miniature detector-analyzer of electrons and protons MiRA_ep.

*Keywords*: radiation belt, electron, nanosatellite, silicon detector, organic scintillator, Brazilian Magnetic Anomaly, inclination of satellite orbit.